\documentclass[]{aastex631}

\usepackage{savesym}
\savesymbol{tablenum}
\usepackage{siunitx}
\restoresymbol{SIX}{tablenum}

\usepackage{amsmath}
\usepackage{mathtools}
\usepackage{subfigure}
\usepackage{siunitx}
\usepackage{textcomp}
\begin{document}

\title{Weak lensing analysis of A115, A2219 and A2261: Detection of galaxy groups and filaments around clusters}

\correspondingauthor{Anirban Dutta}
\email{dutta26@purdue.edu}

\author[0009-0000-1088-4653]{A.Dutta}
\affiliation{Department of Physics and Astronomy \\
Purdue University \\
West Lafayette, Indiana, USA}

\author{J.R.Peterson}
\affiliation{Department of Physics and Astronomy \\
Purdue University \\
West Lafayette, Indiana, USA}

\author{M.Cianfaglione }
\affiliation{Department of Physics and Astronomy \\
University of Bologna \\
Via Gobetti 93/2, 40129 Bologna, Italy}
\affiliation{NAF - Istituto di Radioastronomia\\
Via Gobetti 93/2, 40129 Bologna, Italy}

\author{G.Sembroski}
\affiliation{Department of Physics and Astronomy \\
Purdue University \\
West Lafayette, Indiana, USA}



\begin{abstract}

We present a weak lensing and multi-wavelength analysis of three galaxy clusters: A115, A2219, and A2261. Weak lensing is performed using shape measurements made in short 60s exposure images obtained using WIYN-ODI. Forced measurement is used to measure low Signal to Noise (SNR) sources in individual exposures. We find the weak lensing significance map recovers the galaxy clusters and most galaxy groups in the wide 40$'$ $
\times$ 40$'$ field. Significant parts of the filamentary structures over this field, as indicated by the galaxy number density map, were also successfully recovered in lensing significance maps. We find the amount of structure recovery depends on both the depth and average seeing of the images. In particular, we detect a $>$ 9 Mpc long structure that contains the cluster A2219. We compare our weak lensing maps with Chandra, XMM, and LOFAR observations and find that A115 and A2219 show clear signs of ongoing mergers. In particular, we find a significant separation of hot ICM and the weak lensing contours in A115. On the other hand, while A2261 appears relaxed, based on radio and X-ray analysis, we find that it is likely interacting with a structure 700 kpc SW of the main cluster. We also successfully recovered mass structures in two regions around A2261 indicated by diffuse X-ray emission in XMM images.

\end{abstract}



\section{Introduction} 
Measuring the mass and abundance of galaxy clusters and galaxy groups is fundamental to cosmology \citep{Holder_2001,Cunha_2009}. There exist several methods of measuring the mass of these large-scale structures \citep{sadat_1997}. Lensing is one of the most powerful methods since it directly probes the gravitational potential \citep{tyson_1990, ksb_1993,Schneider_2006}. Several excellent weak lensing studies of galaxy clusters have been performed \citep{okabe_wl_subaru, clash_survey,von_der_Linden_2014, Umetsu_2020, Fu_2024}. Recent surveys such as the Kilo-Degree Survey (KiDS), Dark Energy Survey (DES) and Subaru Hyper Suprime-Cam SSP Survey have produced extremely accurate weak lensing shear measurements by correcting for the major sources of biases \citep{Jarvis_2016, Madelbaum_subaru, Gatti_2021, Kids_2021, Li_2022}. However, one of the long-standing challenges of weak lensing has been the detection of lower mass structures such as galaxy groups and filamentary structures \citep{dietrich2012filament,Maturi_2013}.  These structures are also difficult to detect in other wavelengths such as X-ray or radio because the gas is diffuse, lower in temperature (10$^5$–10$^7$ K), and if of lower density \citep{Tanimura_2020,reiprich_2021,Hoang_2023}. The ability to detect and study these structures is important for our understanding of the Universe, as most of the Universe’s baryonic matter is thought to reside in the Warm-Hot Intergalactic Medium (WHIM), primarily within the filaments of the cosmic web  \citep{simulatons_filaments}. This is also important in our understanding of galaxy clusters since they primarily grow via mergers and accretion along the filamentary structures. Additionally, the cosmic web can be used to understand the nature of dark matter, dark energy, and the influence of the environment on the formation and evolution of galaxies \citep{Libeskind_2017}. These structures are much lower in mass compared to galaxy clusters and hence produce a proportionally lower weak lensing which is challenging to detect. This is further complicated by the subtle shape changes induced by systematics and random noise from a variety of sources. While stacking such images reduces the systematics, it also complicates the PSF and broadly represents a loss of information. It has been proposed that measuring sources in individual exposures would allow for more precise weak lensing measurements compared to shape measurements made in the coadd \citep{tyson_2008, Jee_2011,Mandelbaum_2018}. However, the long-standing challenge was measuring faint sources in the individual exposures. Most shear measurement algorithms start failing at Signal to Noise Ratio (SNR) $\leq$ 10 \citep{Kitching_2012, Kacprzak_2012}. Most sources in individual exposures are expected to have an SNR much lower than 10. A recently proposed generalization of forced photometry solved this issue \citep{dutta_2024a}. It uses guess values of flux, shapes, and centroids derived from the stacked image to then obtain statistically truer values of flux, shapes, and centroids. Using this method, \cite{dutta_2024b} showed that using shape measurements made in individual exposures, one can recover all galaxy groups and most of the filamentary structures in a wide field around the galaxy cluster A2390. They were also able to calibrate this method with simulations to recover the mass of A2390, consistent with existing estimates.

In this paper, we use the same methods to study galaxy clusters A115, A2219, and A2261. These clusters were selected because of their visibility during imaging sessions, deep X-ray images, and their relative proximity. The cluster A115 is a clearly merging cluster at z = 0.18 with two major sub-clusters, one in the north and the other in the south. This is very prominent in the X-ray images \citep{Shibata_1999}. A2261 is a cool core relaxed cluster at z = 0.22 that appears to have a very symmetric and circular profile in X-ray \citep{Allen_2000}. However, it shows signs of a smaller cluster/group falling into the main cluster. A2219 is located at z = 0.22 and appears to be in the process of multiple mergers and smaller accretions from different directions \citep{Canning_2017}. Several X-ray shocks have been detected in this system. We find that for all clusters, using measurements in individual exposures, we can recover the main cluster, a majority of the smaller galaxy groups, and some portions of the filamentary structure around the cluster. The methods followed in this paper are almost identical to \cite{dutta_2024b}. We have therefore describe the methodology briefly.

\section{Observational Data and Processing Pipeline} 
\subsection{Observation with WIYN-ODI}
\begin{figure}[ht!]
    \centering
    \subfigure[]{\includegraphics[width=0.3\textwidth]{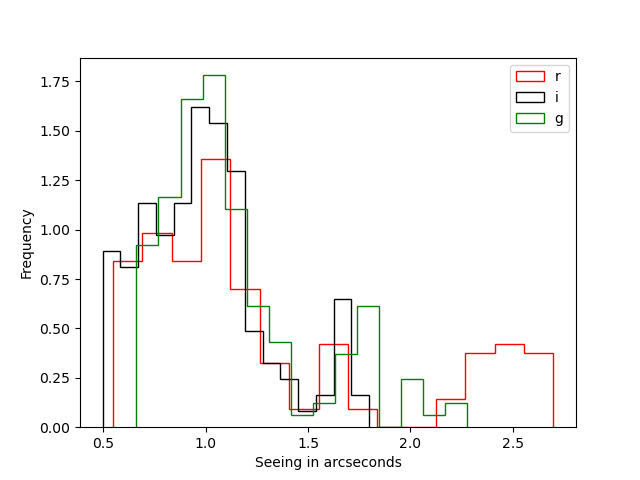}} 
    \subfigure[]{\includegraphics[width=0.3\textwidth]{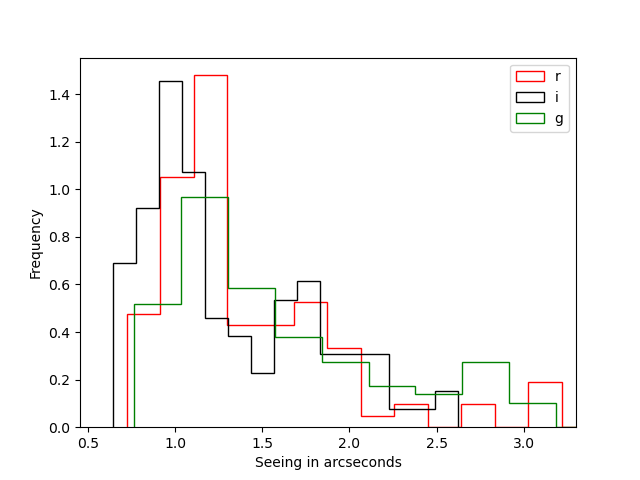}} 
    \subfigure[]{\includegraphics[width=0.3\textwidth]{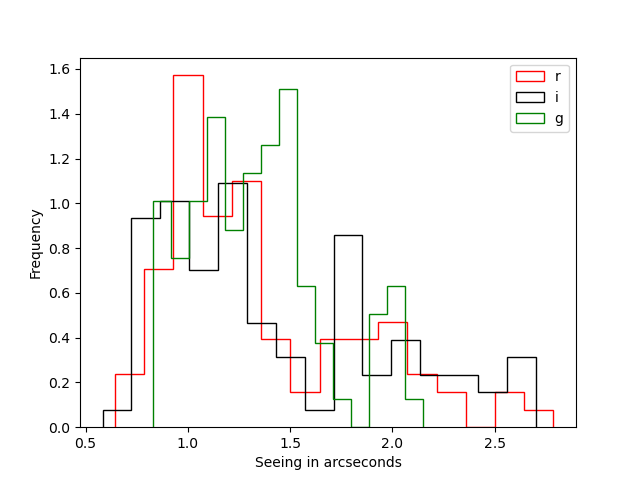}}
    
    \caption{(a) Normalized seeing histogram for g, r and i filters of A115. (b) and (c) show the same for A2219 and A2261} 
    \label{fig:seeing_hist}
\end{figure}

We used the One Degree Imager (ODI) instrument on WIYN, a 3.5m telescope situated in Kitt Peak, Arizona to obtain data for this analysis. The ODI instrument consists of 30 Orthogonal Charge Transfer CCDs (OTAs) arranged in a 5 x 6 pattern. Each OTA is made up of 8 x 8 pixel cell. The image obtained has a pixel scale of 0.11''/pixel. The images were obtained in 5 separate photometric filters: u, g, r, i, and z. This was done to measure photometric redshift. The data for A115 was obtained in the Fall semesters from 2017 to 2023. A2261 and A2219 were imaged in the Spring semesters from 2017 to 2024. We lost a significant portion of the time in the Spring semesters due to weather and telescope issues/downtime. This resulted in a significantly worse quality of data. The number of exposures obtained in each year for each of the three clusters is shown in Table \ref{table:raw_exposure_table}. The overall seeing ranges from excellent (0.6'') to extremely poor (3''). The normalized seeing histogram for each galaxy cluster is shown in Figure \ref{fig:seeing_hist}. 
\begin{table}[h!]
\centering
\begin{tabular}{c c c c } 
 \hline
 Year & Abell 115 & Abell 2219 & Abell 2261  \\ [0.5ex] 
 \hline\hline
 2017   &  42, 38, 42, 42, 44 & - & -\\
 2018   &  26, 26, 26, 24, 25 & - & 0, 9, 9, 9, 9\\
 2019   &  36, 34, 36, 32, 15 & 39, 36, 41, 35, 33 & - \\
 2020   &  43, 36, 36, 35, 25 & - & -\\
 2021   &  18, 17, 9, 9, 9 & 54, 54, 54, 54, 54 & 54, 54, 53, 54, 70 \\
 2022   &  - & - & -\\
 2023   &  - & 10, 18, 18, 18, 9 & 18, 18, 27, 21, 27\\
 2024   &  - & - & 18, 9, 17, 9, 9 \\[1ex] 
 \hline
\end{tabular}
\caption{The number of exposures obtained in each filter for each cluster in each year. The five numbers for each cluster in each year correspond to the number of exposures obtained in filters u, g, r, i, and z respectively.  All exposures are 60s long. Instances, where no data could be obtained due to weather or telescope issues, are shown as "-". A115 was imaged primarily in the Fall semesters and A2261 and A2219 were imaged primarily in the Spring or Summer semesters.}
\label{table:raw_exposure_table}
\end{table}

\subsection{Calibration and co-adding}
We have described the calibration and stacking process in detail in \cite{dutta_2024b}. We mention it here briefly. The initial calibration is performed by the default QuickReduce pipeline \citep{kotulla2013quickreduce}. QuickReduce corrects for all known systematic such as dark subtraction, bias correction, and flat fielding. It also performs photometric and astrometric calibration based on the catalogs of the Sloan Digital Sky Survey \citep{gunn1994sloan}, Pan-STARRS \citep{Pan_stars} and DES \citep{dark2005dark}. Inspection of the calibrated images revealed correlated noise, amplifier glow, non-Gaussian background, and artifacts arising from cross-talk. Correlated noise leads to the detection of several faint sources in the stacked image. Hence, we use the following weight scheme, similar to \cite{Annis_2014} in the co-addition process to suppress noisier images. 
\begin{equation}
W = 100  \frac{10^{Z-25}}{(S  \sigma_b)^2}
\end{equation}
where $Z$ is zero point of the image, $W$ is the weight, $S$ is the average seeing of the image, and $\sigma_b$ is the background variance calculated using the k=3 sigma clipping algorithm in astropy \citep{Astropy1}. This weighting scheme weighs the good seeing images more while reducing contribution from non-Gaussian backgrounds. This allows for better shape measurement in the co-added image. To reject very low pixel values arising from cross-talk we find the lower of 6 standard deviations from the sigma-clipped median background and zero. We reject all pixels lower than this limit. We also rejected the pixel cells using the cuts in the mean vs variance space identical to \cite{dutta_2024b}. It is necessary to use this cut to reject pixel cells that show severe degradation over the course of 7 years of data acquisition. The weighting scheme used is unable to effectively de-weight the severe artifacts in these pixel cells. Before stacking, the images were also visually inspected for windshake, satellite trails, or any other imperfections and flagged. All such flagged images were rejected from the co-addition process.
Coaddition/stacking was performed for each of the five photometric filters using SWarp \citep{swarp}. The stacked i and r band images we co-added further. Source detection is done in the i+r coadded image using Sextractor \citep{sextractor}. The parameters used for SWarp and Sextractor are identical to \cite{dutta_2024b}. In order to limit spurious sources from noisy regions, Sextractor was also run with a detection threshold of 1.5$\sigma$. These regions were visually identified and within these regions, sources detected with the higher threshold were only considered in our analysis. Some small portions of the stacked images were found to be completely unusable due to chip-related artifacts or reflected light from the telescope. These regions were masked out. The overall area of such regions is less than 1 percent of the total image area. For Abell 115, Abell 2219, and Abell 2261 we detect 43419, 50649, and 49615 sources. These correspond to a source density of 25.6, 26.3, and 25.8 sources/arcminute$^2$ respectively. 

\subsection{Source measurement and classification}
The detected sources are measured using an elliptical Gaussian moment matching algorithm developed by \cite{Bernstein_2002}. An elliptical Gaussian is fit to the source and this is used as weight to measure the second moments. The ellipticity is defined as 
\begin{equation}
	e_1 = \frac{Q_{11} - Q_{22}}{ Q_{11} + Q_{22}}  \quad \quad \quad   e_2 = \frac{2Q_{12}}{ Q_{11} + Q_{22}}     
\end{equation}
where $Q_{11}$ the weighted second moment along x axis, $Q_{22}$ the second weighted moment along y axis and $Q_{12}$ is the second weighted moment along y=x line. The moment-matching algorithm also measures the flux, size, and centroid of the sources. It was found that the errors in photon counts, centroid, size, and ellipticity when using this algorithm can be written as a function of photon counts, background, and size of the source being measured. It was also found that this algorithm can be used for forced measurement, a generalization of forced photometry. Forced measurement is used to measure extremely faint sources of SNR $\sim$ 1 or lower. The basic principle is to use the flux and shape measurements of the source obtained from the coadd to construct a guess flux and guess shape. This is then used to run the moment-matching algorithm for one iteration to obtain measurements that are statistically truer than the guess values. For a detailed description and characterization of this method see \cite{dutta_2024a}. The process is described in detail in Section 2.5. The detected sources are first measured in the i+r coadded image followed by measurements made in the stacked images of each filter. To make the cutouts an initial cutout size is estimated which is defined as 1.25 times the SExtractor size. This cutout size is then refined to reduce light contamination from nearby sources thereby allowing for better measurement. The flux measurements in each filter is used to measure photometric redshift using EAZY \citep{Brammer_2008}. 

\begin{figure}[ht!]
    \centering
    \subfigure[]{\includegraphics[width=0.3\textwidth]{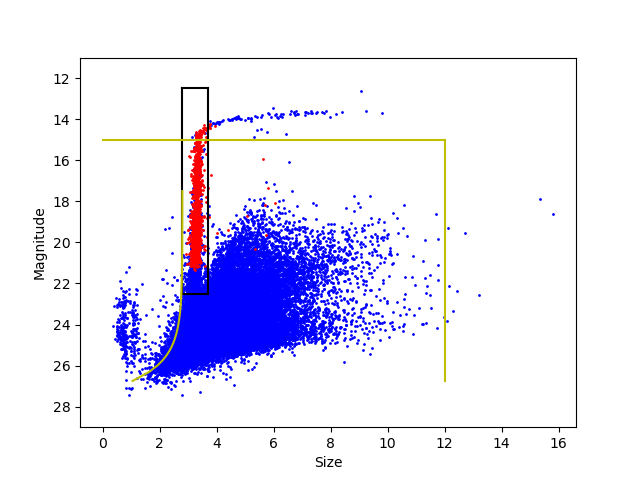}} 
    \subfigure[]{\includegraphics[width=0.3\textwidth]{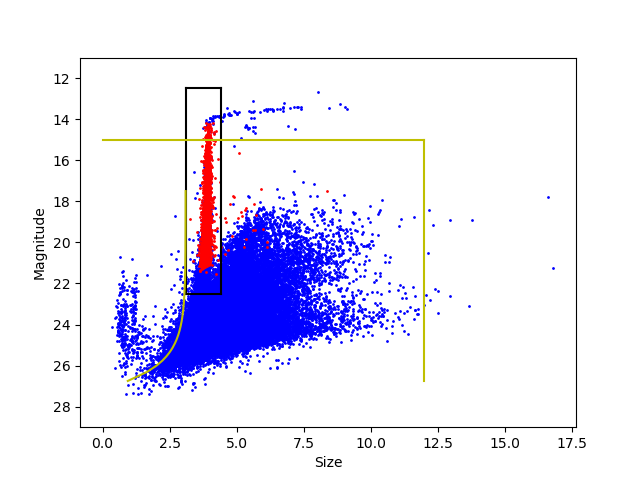}} 
    \subfigure[]{\includegraphics[width=0.3\textwidth]{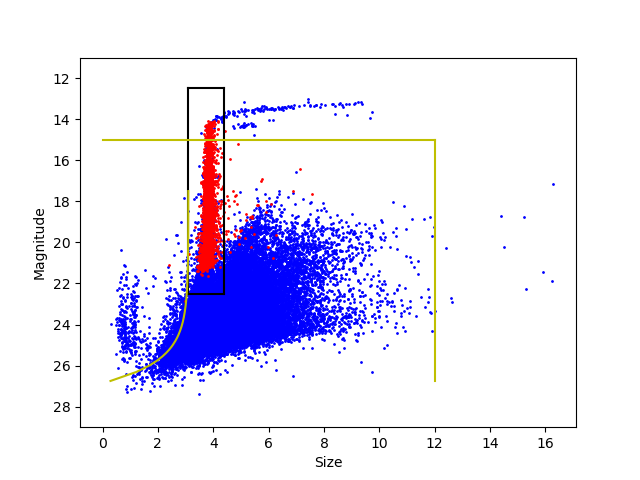}}
    
    \caption{(a) Magnitude brightness in the i+r coadd vs size for all sources detected in the i+r coadd of Abell 115. The points in red show sources cross-matched with Gaia. The sources in the black box are rejected from the weak lensing analysis. The yellow line on the top shows the limit at which brighter fatter effect becomes important. The yellow line on the right shows sources that are rejected from the weak lensing analysis. The yellow line on the left shows sources that are 3 standard deviations smaller than the PSF. Sources to the left of this line are rejected from our analysis. (b) and (c) shows the same for Abell 2219 and Abell 2261 respectively. } 
    \label{fig:flux_vs_size}
\end{figure}

Stars are useful for determining the Point Spread Function (PSF). However, they must be rejected from weak lensing analysis. We identify stars by cross-matching our sources with sources from Gaia EDR3 \citep{gaia_edr3}. The sources that are successfully cross-matched and fainter than 15 magnitude is used for PSF interpolation. Stars brighter than magnitude brightness 15 show brighter fatter effect. PSF interpolation is performed by considering the 15 nearest stars from any given location. The inverse distance weighted $\sigma^2_{xx}$ , $\sigma^2_{yy}$ and $\sigma^2_{xy}$ is defined as the interpolated values at that point. Cross-matching sources with Gaia is unable to identify all stars in our image since EDR3 only goes to a depth of 21 magnitude in the g band while our images have a depth of 26 magnitude. Hence, cuts are performed in the magnitude vs size space to remove the point sources not identified by Gaia. The graph of size vs magnitude brightness in the i+r coadd of the three galaxy clusters is shown in Figure \ref{fig:flux_vs_size}. The red points show the sources cross-matched with Gaia. The sources inside the black rectangle are identified as stars and are excluded from the weak lensing analysis. The limits of the black box were determined after visual inspection. For A115, the left and right limits of the black box are from 2.8 pixels to 3.7 pixels respectively. For A2219 the limits are 3.1 pixels and 4.4 pixels. For the dataset containing A2261 the limits for the black box are 3.1 pixels and 4.4 pixels. The yellow line on the right is drawn at 12 pixels and shows the size above which we reject sources. The yellow line on the top is drawn at magnitude 15. Any source brighter than this limit is rejected because beyond this limit brighter fatter effect becomes important. The yellow line on the left is drawn along the left edge of the black box, after subtracting 3 times the Poisson error in size. Hence, as the flux decreases, Poisson error increases causing the line to move to the left. The Poisson error in the size of a source is given by 
\begin{equation}
\epsilon_{size} =\sqrt{\frac{1}{2} \frac{ S^2}{ N} \left(1  + K \right) }
\end{equation}
where $S$ is the size, $N$ is the number of photon counts and $K$ for elliptical Gaussian case is $4\pi S^2B / N$. The background level is given by $B$. Size of a source is defined as 
\begin{equation}
S =\sqrt{\sigma^2_{xx} + \sigma^2_{yy}}
\end{equation}

\subsection{PSF Correction}
The weak lensing signal is typically at a few percent level, comparable to the level of PSF variation across the image. Hence, PSF correction is necessary to recover the weak lensing signal. It was found about 35 \% of sources are measured to be smaller than PSF. Traditionally sources smaller than the size of the PSF have been discarded from weak lensing analysis. This was done because sources smaller than PSF, by definition, are unphysical. The existing methods of PSF correction are unable to process these sources. However, it was found that including these sources can improve the recovery of smaller galaxy groups and filamentary structures as shown in \cite{dutta_2024b}. Thus, we use the same method and perform Monte Carlo PSF correction. The idea behind this method is that the sources that are measured to be smaller than PSF are simply a consequence of statistical fluctuation and large error bars. Knowing the error in the second moments of the PSF and the source, one can perform Monte Carlo sampling to estimate the true shape. Assuming both the source and PSF are approximately described by elliptical Gaussian, we can write
\begin{equation}
\sigma^2_{xx} (true) = \sigma^2_{xx}(measured) \pm \sqrt{ \frac{S^4 }{ N} (1 + K) } -  \sigma^2_{xx} (PSF) \pm \epsilon(xx)_{PSF}
\label{equn:mc_psf_corr}
\end{equation}
where $\sigma^2_{xx}(true)$ is the true value of a source, $\sigma^2_{xx}(measured)$ is the measured value, $\sigma^2_{xx}(PSF)$ is the interpolated value for PSF and $\epsilon(xx)_{PSF}$ is the error in PSF value. The expression $\sqrt{ S^4 (1 + K)/N }$ is the Poisson error in $\sigma^2_{xx}(measured)$, where $S$ is measured size, $N$ is measured photon counts and $K$ for elliptical Gaussian case is $4\pi S^2B / N$. $B$ is the background level. Similar equation is used to determine $\sigma^2_{xy}(true)$ and $\sigma^2_{yy}(true)$
To determine $\epsilon(xx)_{PSF}$, the error from the PSF variation across the image is added in quadrature to the error arising from the Poisson noise of the stars used to interpolate PSF. After all the terms on the right-hand side of Equation \ref{equn:mc_psf_corr} have been determined, we solve it using the Monte Carlo method. We draw 30,000 samples of $\sigma^2_{xx}$ , $\sigma^2_{yy}$ and $\sigma^2_{xy}$ where the uncertainty in Equation \ref{equn:mc_psf_corr} (second and fourth terms) are drawn from a random normal distribution whose width is equal to the uncertainties. We only consider those samples where both $\sigma^2_{xx}$ and $\sigma^2_{yy}$ are greater than 0 and the absolute value of $e_2$ is less than 1. These conditions were found to be optimal in shear recovery. In order to ensure our sampling estimates are robust, we reject any sources where less than 50 samples successfully pass our cuts. We calculate the median of all samples that pass our cuts and call them the true value of $\sigma^2_{xx}$ , $\sigma^2_{yy}$ and $\sigma^2_{xy}$.

\subsection{Measurement in Individual Exposures}
It has been proposed \citep{tyson_2008, Jee_2011, Mandelbaum_2018} and demonstrated \citep{dutta_2024b} that measuring shear in individual exposures improves shear recovery. Hence, we perform shape measurements in individual exposures. We decided to use individual exposures in the filters g, r, and i for two reasons. First, we generally have the most amount of data and good quality seeing in these bands. And secondly, most of the sources are brightest in these bands, especially in the i and r filters. In the first step, we measure all the bright sources including the stars in the coadd. For each source, we estimate the SNR in each individual exposure. This is obtained by using guess flux ($N_{expected}$) and guess size. 
\begin{equation}
N_{expected} = N_{coadd} \times \left \langle \frac{N_{image}(star)}{N_{coadd} (star)}\right \rangle 
\end{equation}
\begin{equation}
\sigma^2_{image} = \sigma^2_{coadd} (true) + \sigma^2_{image}(PSF)
\end{equation}
where $\sigma^2_{coadd}(true)$ is obtained after using the Monte Carlo PSF correction method described above to solve Equation \ref{equn:mc_psf_corr} in the i+r co-add. $\sigma^2_{image}(PSF)$ is the interpolated PSF at the location in the given exposure. The quantity $\langle N_{image}(star)/N_{coadd} (star) \rangle$ is the ratio of photon counts in the image to that of the stacked image in a given filter. The angle bracket denotes the median value of the 10 nearest stars are used. The signal-to-noise is defined as 
\begin{equation}
SNR = \frac{N}{\sqrt{N + 4AB}}
\end{equation}
where $N$ is the photon counts from the source, $A$ is the area and $B$ is the background. We use the quantities calculated in Equation 5 and 6 to calculate the estimated SNR. If the estimated SNR of a source in an individual exposure is greater than 15, we run the moment-matching algorithm until convergence. If the SNR is less than 15 or the convergence fails, we turn to forced measurement. 

A vast majority of the sources detected in the stacked image is extremely faint in the individual exposures (SNR $\leq$ 5). Under these conditions, the traditional moment-matching algorithm is unable to converge to a stable solution. Hence, we use forced measurement \citep{dutta_2024a}. Forced measurement is a generalization of the forced photometry method to measure ultra-low S/N sources. First, the negative pixel values after background subtraction are corrected to ensure all pixels are positive. In the second step, guess the values of flux, centroid and $\sigma^2_{xx}$ , $\sigma^2_{yy}$ and $\sigma^2_{xy}$ constructed from the stacked image are used to perform a single iteration of the moment matching algorithm. The values thus obtained are corrected for bias in flux and size. The bias correction step is necessary because of the first step where all negative pixel values were converted to positive values. Bias correction is done using the look-up table of \cite{dutta_2024a}. This method produces values of flux, centroid, and $\sigma^2_{xx}$, $\sigma^2_{yy}$ and $\sigma^2_{xy}$ that are statistically closer to the true values. 

After forced measurement, we perform PSF correction using the Monte Carlo method described above. We use a slightly different equation compared to Equation \ref{equn:mc_psf_corr}
\begin{equation}
\sigma^2_{xx} (true) = \sigma^2_{xx}(measured) \pm p_{\sigma^2_{xx}}\sqrt{ \frac{S^4 }{ N} (1 + K) } -  \sigma^2_{xx} (PSF) \pm \epsilon(xx)_{PSF}
\label{equn:mc_psf_corr_sf}
\end{equation}
where $\sigma^2_{xx}(measured)$ is the values measured by forced measurement in individual exposures, S is the size, N is the photon counts, $\sigma^2_{xx} (PSF)$ is the interpolated PSF and $\epsilon(xx)_{PSF}$ is the error in the interpolated PSF. The additional factor $ p_{\sigma^2_{xx}}$ comes from the fact that for forced measurement we perform a single iteration and give the algorithm a good initial guess of the flux, shape, and centroid. If the source being measured is bright (SNR $\geq$ 15), and the moment matching algorithm successfully converges, this factor is set to 1.

Measurements in individual exposures are by definition noisy. Especially problematic are correlated noise and cross-talk effects when measuring very faint sources. These can completely drown the signal from the source and produce incorrect measurements. Hence, we apply several cuts to the individual exposure measurements to reject outliers. We reject any measurement where nan's or negative values of flux have been obtained. We compute the sigma clipped standard deviation of $\sigma^2_{xx}$ and $\sigma^2_{yy}$. Any measurement that yields values that are 3 standard deviations away from the median is rejected. We also reject measurements from individual exposures where $\sigma^2_{xx} (PSF) \geq 30$ and $\sigma^2_{yy} (PSF) \geq 30$. This corresponds to extremely bad seeing and such images are unsuitable for weak lensing. Cases where $\sigma^2_{xx} \leq 1$ and $\sigma^2_{yy} \leq 1$ are also rejected.  If the sigma clipped standard deviation of $\sigma^2_{xx}(PSF)$ or $\sigma^2_{yy}(PSF)$ across the image is greater than 3 pixels, we reject all measurements from those images. It was found such variations are unusually large and typically indicate issue with the images. The remaining measurements are combined using inverse error squared as a weight to obtain the final shape measurements. 
\begin{equation}
\sigma^2_{xx} = {\sum_{i}\sigma^2_{xx,i} W(i) }
\end{equation}
where $\sigma^2_{xx,i}$ is the PSF corrected measurement in the i-th frame and the weight of the i-th image is $W(i)$. The total error is calculated by adding the Poisson error of the source and the PSF error in quadrature. This error is inverse squared to obtain the weight. A similar process is followed for $\sigma^2_{xy}$ and $\sigma^2_{yy}$

To recover shear from the measurements made in individual exposures, we follow a method similar to GREAT3 \citep{Mandelbaum_2014}. 
\begin{equation}
g_t \approx \gamma_t = \frac{\sum_{i=1} e_{t,i} w_{a,i} }{2 (1- \langle e^2 \rangle)}
\label{eqn:ellip_to_shear}
\end{equation}
where $\gamma_t$ is the tangential ellipticity, $e_t$ is the tangential ellipticity of a source and $g_t$ is the reduced tangential shear. $\langle e^2 \rangle$ is the average value of ellipticity squared. The weight $w_{a,i}$ is defined as
\begin{equation}
w_{a,i} =  \frac{ \epsilon^{-2}_{e,i} }{\sum_{i=1}\epsilon^{-2}_{e,i}}
\end{equation}
where $\epsilon_{e}$ is the error in ellipticity. The error in ellipticity is computed by propagating the error in $\sigma^2_{xx}$, $\sigma^2_{xy}$ and $\sigma^2_{yy}$. We use the form presented in \cite{dutta_2024b}. We also define
\begin{equation}
\langle e^2 \rangle =  \sum_{i=1} e_i^2 w_{b,i}
\end{equation}
The weights $w_{b,i}$ is defined as 
\begin{equation}
w_{b,i} =  \frac{ \epsilon^{-1}_{e,i} }{\sum_{i=1}\epsilon^{-1}_{e,i}}
\end{equation}
where $\epsilon_{e}$ is the error in ellipticity. This method of extracting shear was found to be optimal after testing on simulated images from PhoSim \citep{Peterson_2015, Burke_2019,Peterson_2019, Peterson_2020, peterson_2024}. 

\subsection{Aperture Mass Maps}
Aperture mass maps are useful in detecting mass peaks. It measures the average tangential shear around a given point using a given weight. Different weight functions, depending on the structures being detected, have been proposed in the literature \citep{Maturi_2013, McClearly_2020}. Some are optimized to find filamentary structures, while others are optimized to detect NFW profiles. We use a Gaussian weight function. Aperture mass $M_{ap}$ and the corresponding error $\sigma_{M_{ap}}$ are defined as 
\begin{equation}
M_{ap} = \sum_{i}w(|\theta| - |\theta_i|) w_p(i) g_{i,t}
\label{equn:m_ap}
\end{equation}
\begin{equation}
\sigma_{M_{ap}} = \sqrt{\sum_{i} w^2(|\theta| - |\theta_i|) w_p(i)}
\end{equation}
where $g_{i,t}$ is the tangential shear of the i-th galaxy and $w_p(i)$ is the weight function which is inverse Poisson error squared in ellipticity, and the overall weight function is denoted by $w(|\theta|)$. The overall weight function is a Gaussian in our case. The size of the Gaussian is typically a few arcminutes. It has been proposed by \cite{McClearly_2020} that aperture mass statistics are optimized for detecting structures that are best matched with the weight function. Thus it behaves similar to a matched filter. In our case, it was found that using a wider weight function leads to the recovery of a wider structure. On the other hand, using a narrower Gaussian as weight leads to recovery of the smaller mass structures, albeit with larger errors. As a result, when using a narrower Gaussian to detect shear from a galaxy cluster, only the central overdensity can be detected prominently. The overall cluster halo is not recovered with a high significance. Hence, we use three different weight widths (4.5', 3.6', 2.3') and average over the significance maps. The significance maps are defined as $M_{ap}/\sigma^2_{M_{ap}}$. This allows us to capture both the larger halo and the smaller central overdensity of the galaxy clusters. It was also found that $M_{ap}/\sigma^2_{M_{ap}}$ map produces less noise and captures more structure than the traditionally used $M_{ap}/\sigma_{M_{ap}}$. 

It was found that the results of the significance map depend on the number of co-adds. The cause of this was determined to be inaccurate aperture mass error maps. This is expected because our methods only take into account the Poisson error and the variation in PSF across an image. However, in individual exposures, the signal from the source is extremely faint (SNR $\sim$ 1) and is comparable to the noise arising from defects in the CCDs. These include correlated noise, non-Gaussian background, and effects of crosstalk. Hence, the errors are underestimated, and as the number of frames increases, so does the level of underestimated error. This was found to be true in our case as well. To solve this, we consider all the pixels in the significance maps and find the sigma-clipped standard deviation. It was found drawing contours at 0.6, 1.5, and 2.5 standard deviations yielded good results. Hence, these levels were adopted. The results obtained using this method is only slightly sensitive to the error map. 

\subsection{X-ray images}
The Chandra X-ray images are processed using CIAO \citep{ciao}. We searched for existing Chandra data on the Chandra Data Archive. Available observations were downloaded and merged using the "merge\_obs" command. For Abell 115 the observation IDs of the datasets merged are 3233, 13458, 13459, 15578, 15581. This leads to a total exposure time of 331ks. For Abell 2219, the observation IDs used are 896, 7892, 21968, 21967, 21966, 20952, 20951, 20785, 20589, 20588, 14451, 14431, 14356, 14355, 13988 with a total exposure time of 495 ks. We use the exposure-corrected broad band image in our figures. The "merge\_obs" command outputs this file as broad\_flux.img. The unit of the output image is counts/cm$^2$/s. For Abell 2261, we used the XMM image instead of the Chandra image because of its more uniform coverage and wider field of view, which allows us to identify at least two smaller groups and clusters in the field. We used the XMM images that were released as a part of LOFAR DR2. The data processing is described in \cite{botteon_2022}.

\subsection{LOFAR Images}
We re-analyzed the data from the Second Data Release of the LOFAR Two Meter Sky Survey \citep{botteon_2022} of the clusters Abell 2219 and Abell 2261. The initial images are obtained by the combination of three 8 hour pointings per target, hence reaching a total observation time of 24 hours per cluster. In order to detect the radio halos present in these clusters, we followed a source subtraction approach tuned ``ad-hoc" for these targets. Specifically, we started from 27" resolution images (corresponding to a taper of 100 kpc) from which we subtracted from the visibilities all the sources present on scales smaller than $r_{500}$, which means smaller than 1490 kpc (408") for Abell 2219 and 1302 (362") for Abell 2261, and then produced new images at the same resolution. This allowed us to properly remove the contribution from the powerful radio-galaxies embedded in the cluster and to increase the sensitivity towards the presence of radio halos. For A115, we used the images directly from the LOFAR Two Meter Sky Survey. 

\section{Results}
\subsection{Abell 115}
Abell 115 has been studied extensively over the last two decades\citep{Shibata_1999, Chatterjee_2024}. It is located at z=0.19 and seems to be undergoing a major merger \citep{Barrena_2007}. It contains a radio relic almost 1 Mpc long \citep{Govoni_2001, Botteon_2016, Chatterjee_2024}. The galaxy cluster shows two distinct clumps in X-ray \citep{Gutierrez_2005,Hallman_2018}. There has been significant debate in the literature as to which clump is heavier \citep{Kim_2019}. Even weak lensing studies with the same data came to different conclusions and locations of mass centroids \citep{okabe_wl_subaru, Kim_2019}.

\begin{figure}[ht!] 
    \centering
    \subfigure[]
    {\includegraphics[width=0.85\textwidth]{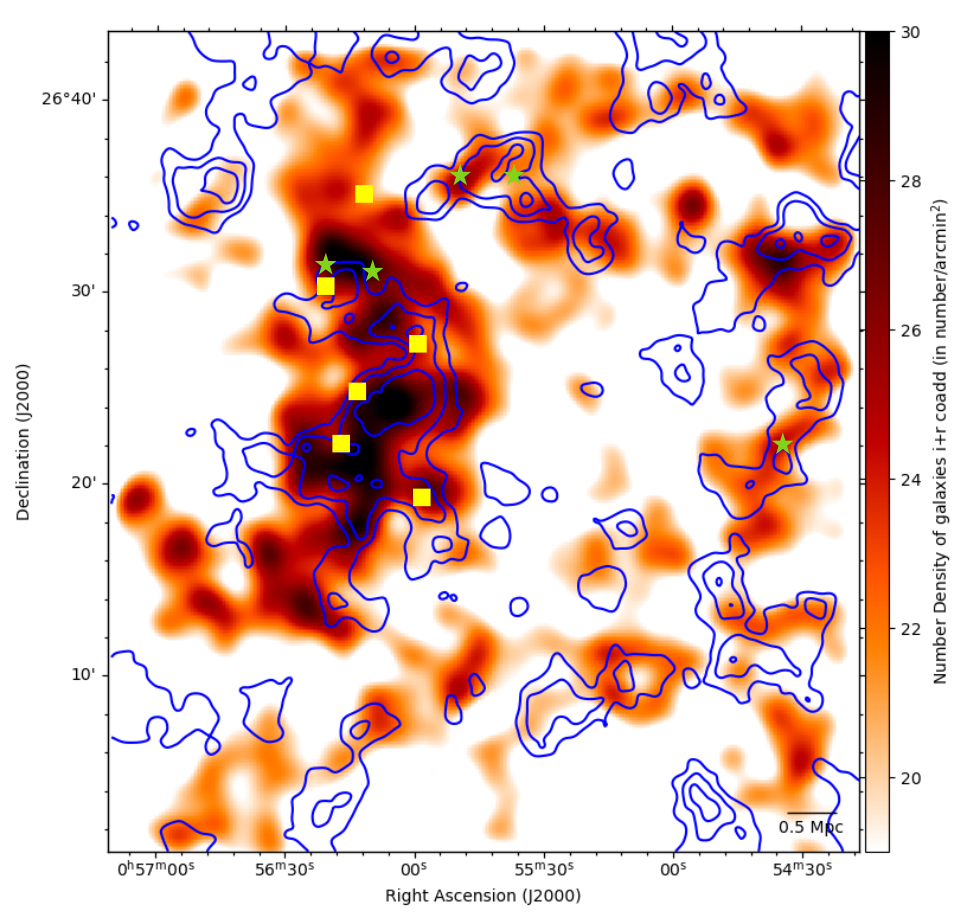}}
    
    \caption{Contours of the E-Mode significance map of A115 overlayed on the galaxy number density map of all sources except stars. The contours are drawn at 0.6, 1.5, and 2.5 standard deviations. The green stars show galaxy clusters and groups identified from NED using optical methods while yellow squares show clusters and groups obtained using X-ray.}
    \label{fig:A115}
\end{figure}

\begin{figure}[ht!] 
    \centering
    \subfigure[]{\includegraphics[width=0.45\textwidth]{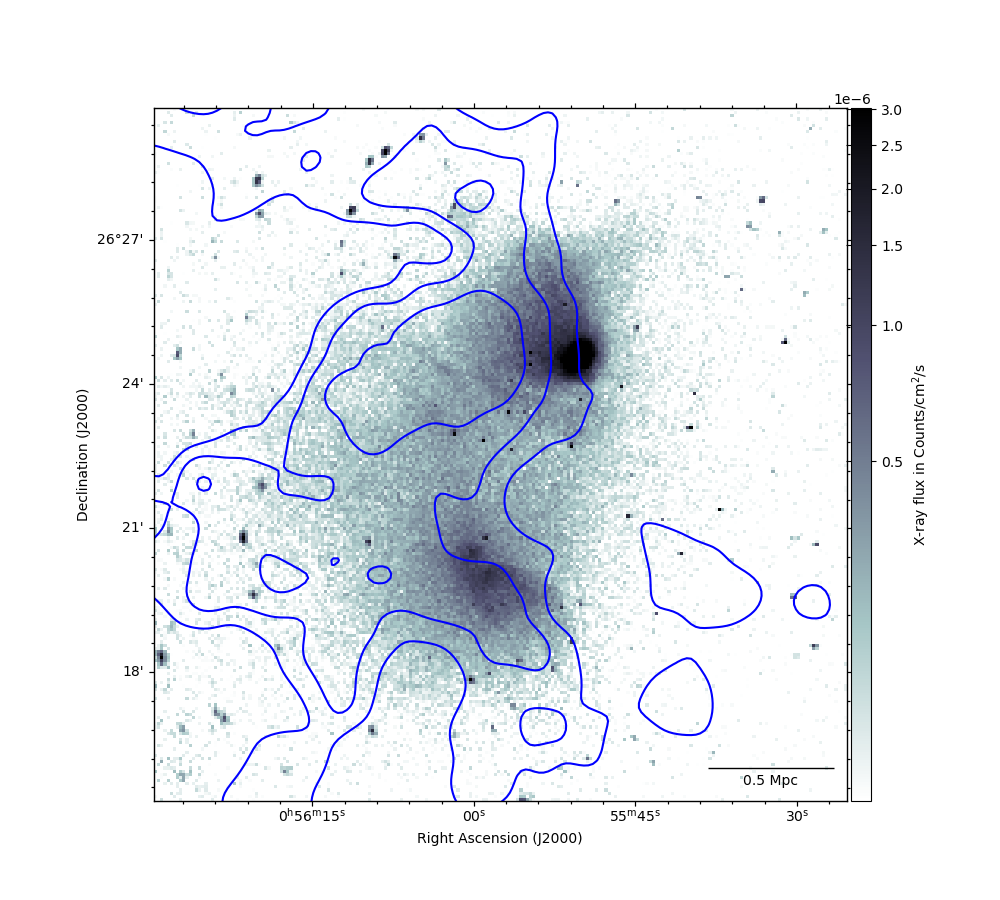}}
    {\includegraphics[width=0.45\textwidth]{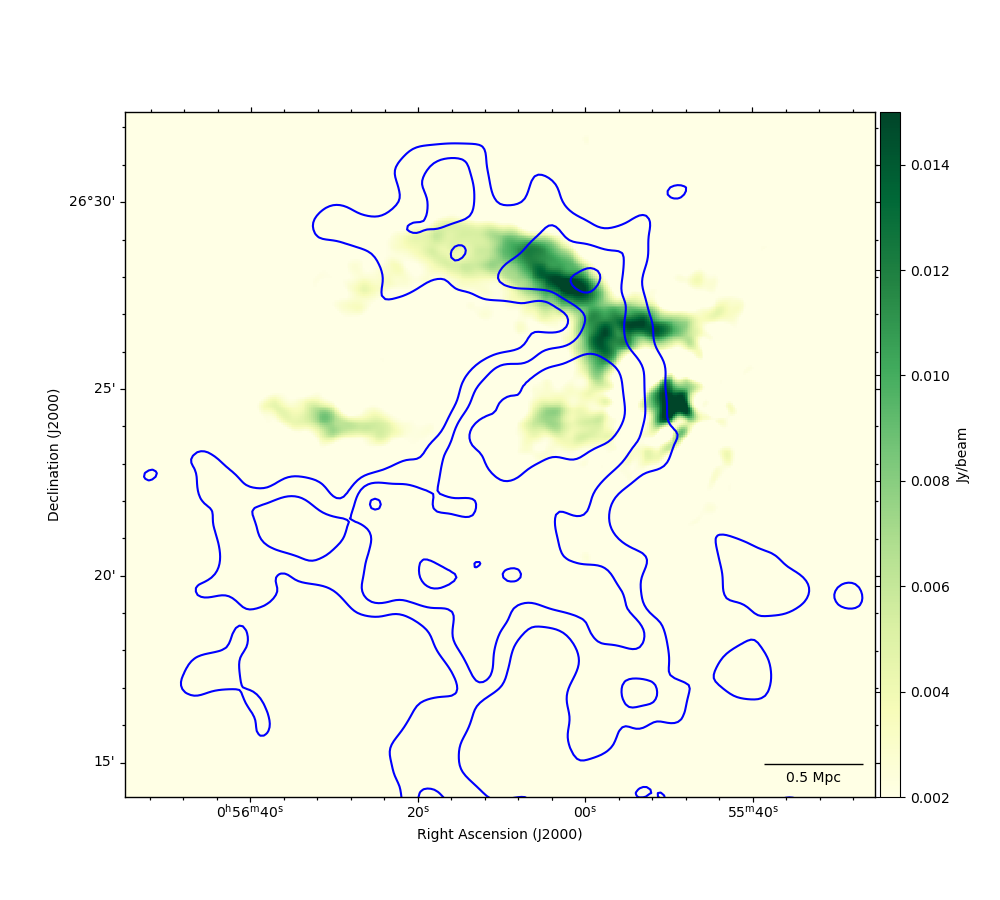}}
    \caption{(a) Contours of the E-Mode significance map of A115 overlayed on the Chandra X-ray image with approximately 361 ks exposure time. (b) The same contours are overlayed on the LOFAR image after point source subtraction with a 50kpc taper. }
    \label{fig:A115_xray_radio}
\end{figure}

In our analysis, we find that our mass map is significantly different from the published literature. This might be because our images are 1-2 magnitudes deeper than the stacked Subaru image \citep{okabe_wl_subaru}. The previous weak lensing studies of A115 have been performed using the Subaru images. Specifically, we find mismatch of the ICM with the lensing contours, while the number density is in agreement with the lensing map. Figure \ref{fig:A115} shows the E-Mode contours overlayed on the galaxy number density map which includes all sources, excluding stars. Stars and point sources are identified using the methods mentioned in Section 2. Contours are drawn at standard deviation values of 0.6, 1.5, and 2.5. The location of galaxy groups and clusters found using NED are shown as symbols. The yellow squares show clusters/groups identified using X-ray techniques \citep{Haines_2018, Giles_2022}. The green stars show groups and clusters identified using optical methods, primarily from the Sloan Digital Sky Survey data \citep{Koester_2007, Estrada_2007,McConnachie_2009,Hao_2010,Wen_2012, Wen_2015, Rozo_2015, Kirkpatrick_2021}. For visual clarity, if both X-ray and optical methods identify a cluster/group at a given location, we only show one symbol. Since this is an actively merging cluster, we only show the galaxy groups and clusters that are at least 4' (0.76 Mpc) from the centroid. Figure \ref{fig:A115_xray_radio}a shows a zoomed-in image of the E-mode contours overlayed on the 331ks Chandra X-ray image of Abell 115. Finally Figure \ref{fig:A115_xray_radio}b shows the E-Mode significance contours overlayed on the radio images from LOFAR with 50kpc subtraction \citep{botteon_2022}. It appears that the radio relic is almost perfectly traces the E-Mode significance contours and galaxy number density. We find that our mass maps are consistent with the number density of galaxies and are able to recover most galaxy groups and clusters reported in the field. At the center of the cluster, we find a separation between DM core and X-ray ICM gas. The DM profile from weak lensing roughly traces the galaxy number density. This is expected in mergers of galaxy clusters where the dark matter and individual galaxies behave approximately collisionless.

Spectroscopic analysis of this cluster suggests that the plane of the merger is close to the plane of the sky i.e. approximately 20 degrees \citep{Barrena_2007}. It is generally accepted, based on the X-ray morphology, that the northern cluster is moving to the right while the lower sub-cluster is moving to the left, orbiting their common center of mass \citep{Hallman_2018,Kim_2019}. The gas from both the clumps is colliding near the center which leads to a higher temperature in that region and is seen in the X-ray temperature maps. An X-ray shock was reported by \cite{Botteon_2016} near the radio relic. However, \cite{Hallman_2018} were not able to replicate this finding and instead found a shock just ahead of the northern clump. It appears from spectroscopic data that the northern sub-cluster is moving into the plane (radially) faster than the southern sub-cluster. Along with the galaxy number density and weak lensing mass maps, this indicates that the northern clump, after moving to the right the maximum amount, has recently started its move back to the center which lies to its left and likely into the plane. The DM, as expected is leading the baryonic matter i.e. the X-ray plasma. On the other hand, the lower clump having moved to the left and into the plane is likely also falling back to the center of mass which is to its right and radially out of the plane. This explains why the farthest clump of the galaxies related to the southern clump has a lower spectroscopic velocity as compared to galaxies related to the northern sub-clump. It also explains the kink seen in the Chandra X-ray image of the lower sub-cluster. The hot ICM gas in the southern clump is also falling into the center of mass which lies to its north and likely out of the plane. In this case, the dark matter is leading the hot gas as expected. However, we note here that the lower sub-clump has a very elongated structure, unlike what would be expected in a merger. Not only is this structure seen in weak lensing, it is also seen in the number density of galaxies and roughly matches the spectroscopic finding of \cite{Barrena_2007}. Simulations of galaxy cluster mergers done by \cite{ZuHone_2011,Kim_2019} do not show features like this. While this is very interesting, a simpler and likely explanation is that a smaller unrelated clump falling in the cluster. Only a deeper spectroscopic investigation can shed more light on this structure. 

\subsection{Abell 2219}
Abell 2219 is a galaxy cluster at z=0.22. It has been studied extensively and appears to be undergoing a major merger \citep{smail_1995,Allen_1998, gray_2000, Coia_2005, Orru_2007,Million_2009,Prokhorov_2011}. The X-ray images show complex morphology and evidence of multiple shocks and discontinuities in the surface brightness profile \citep{Canning_2017}. They hypothesize that Abell 2219 is undergoing a merger along the SE-NW direction. They also find a major infalling clump that is currently very close to the closest crossing. Radio images at 1.4 GHz and 325 MHz from VLA show a diffuse halo indicating turbulence in the ICM \citep{bacchi_2003,Orru_2007}. The spectral index was found to vary significantly throughout the halo. A multi-wavelength analysis of A2219 has been done by \cite{Boschin_2004}. They also find evidence of elongation of the cluster along the SE-NW direction. Weak lensing analysis of this cluster was also performed by \cite{okabe_wl_subaru, von_der_Linden_2014} with deep images obtained from Subaru. While \cite{okabe_wl_subaru} found mass maps that are overall elongated in NE-SW direction at the center, the aperture mass maps of \cite{von_der_Linden_2014} is almost symmetric and circular. 

\begin{figure}[ht!]
    \centering
    
    \subfigure[]{\includegraphics[width=0.85\textwidth]{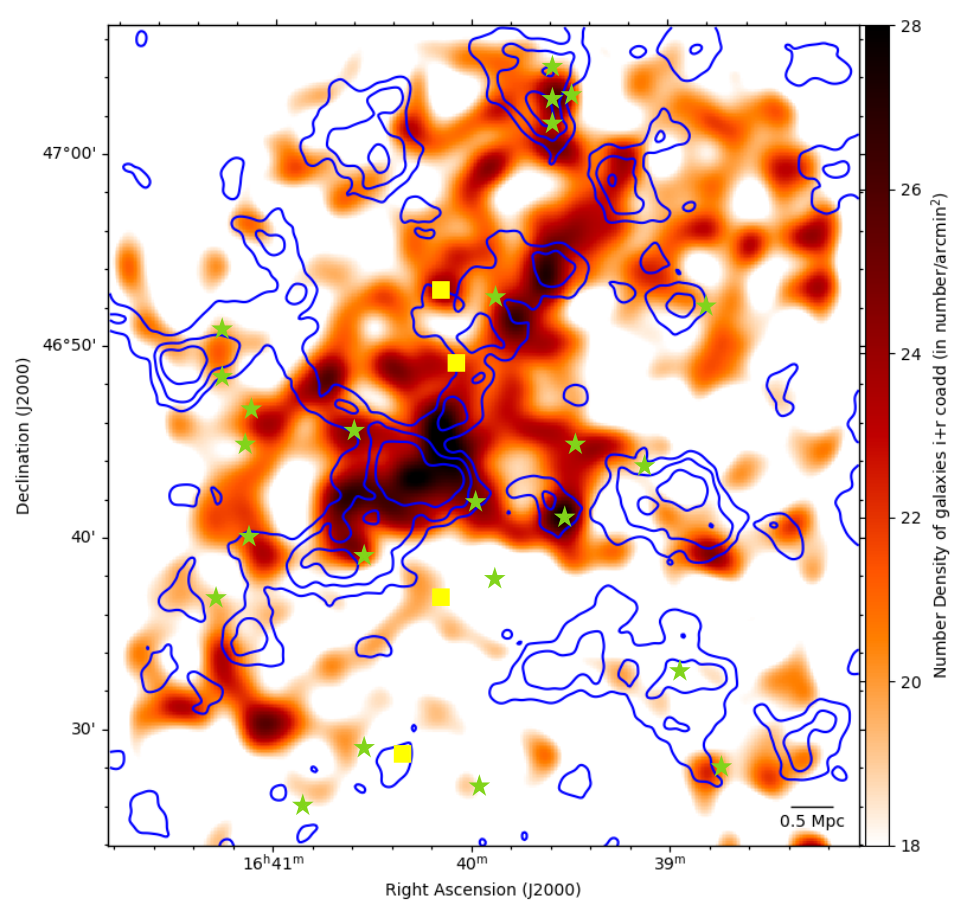}}
    
    \caption{ Contours of the E-Mode significance map of A2219 overlayed on the galaxy number density map of all sources except stars. The contours are drawn at 0.6, 1.5, and 2.5 standard deviations. The green stars show galaxy clusters and groups identified from NED using optical methods while yellow squares show clusters and groups obtained using X-ray. }
    \label{fig:A2219}
\end{figure}

\begin{figure}[ht!]
    \centering
    \subfigure[]{\includegraphics[width=0.45\textwidth]{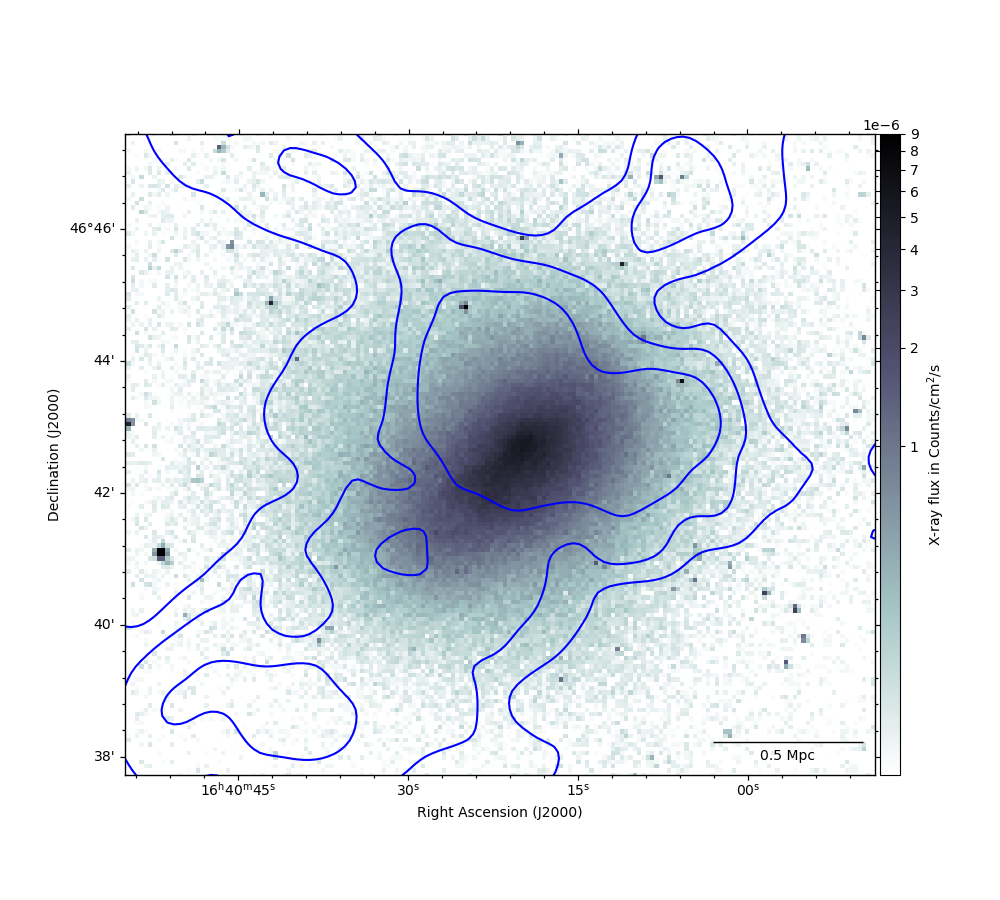}} {\includegraphics[width=0.45\textwidth]{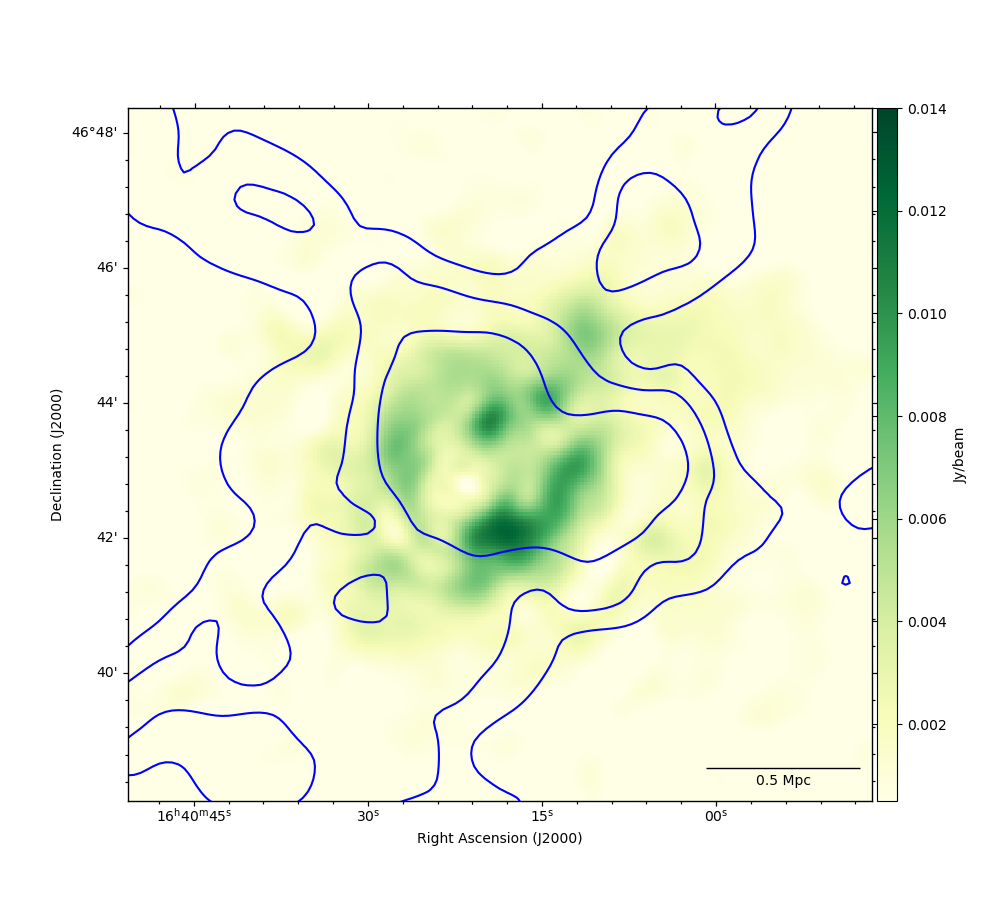}}
    \caption{(a) Contours of the E-Mode significance of A2219 map overlayed on the Chandra X-ray image with approximately 495 ks exposure time. (b) The same contours are overlayed on the LOFAR image.  }
    \label{fig:A2219_xray_radio}
\end{figure}

In our analysis, we find that we successfully recovered the main cluster in the mean E-Mode significance maps. Figure \ref{fig:A2219} shows the E-Mode contours overlayed on the galaxy number density map which includes all sources except stars. Contours are drawn at standard deviation values of 0.6, 1.5, and 2.5. The central part shows an elongation along the NE-SW direction as found by \cite{okabe_wl_subaru}. However, on a larger scale, the elongation is along NW-SE. The galaxy clusters and groups reported on NED and at least 3$'$ from the cluster are shown as green stars (optical methods) or yellow squares (X-ray methods). We find that most of the galaxy groups and clusters reported on NED in our field of view are recovered, albeit with a lower significance. This is not surprising since these structures have masses that are an order of magnitude lower than the cluster mass. Qualitatively, it appears that the recovery of mass structures and filaments is worse than A115 field. This is likely due to 30\% more exposures we have for A115. We also recovered a 9.2 Mpc long filament in the NW-SE direction. It appears that the cluster is embedded inside this filament. The high number density of galaxies in the region also confirms this. Hence, it is likely that the cluster is accreting material from both SE and NW directions.  Previous weak lensing analyses of comparable and higher depth \citep{okabe_wl_subaru, von_der_Linden_2014} have not been able to detect this filament. The X-ray image from Chandra with approximately 495ks of exposure with the E-Mode contours overlayed is shown in Figure \ref{fig:A2219_xray_radio}a. The radio image from LOFAR with the E-Mode contours overlayed is shown in Figure \ref{fig:A2219_xray_radio}b. 

Our results are consistent with the existing studies. It is clear that the cluster exists in a massive $\sim$ 10 Mpc filament which extends in the NW-SE direction and likely has a component perpendicular to the plane of the sky \citep{Boschin_2004,Canning_2017}. It appears from the galaxy number density maps in Figure \ref{fig:A2219} that the cluster is undergoing mergers from multiple directions. This explains multiple surface brightness discontinuities found in the system. The most prominent ones appear to be from SE, NW, and W. The SE substructure detected by \cite{Boschin_2004} about 2' from the center is also detected in out E-Mode significance map. They classified it as a foreground pre-merger clump based on color, spectroscopic, and X-ray observations. We find no evidence opposing this. It was noted in \cite{Orru_2007,Canning_2017} that the spectral index steepens in the NW and W directions, unlike in the SE direction. This could be due to the fact that an active merger is happening from NW and W while the southern part is somewhat inactive at the moment, having completed a major merger. It could also be the case, as suggested by \cite{Canning_2017} that the recent merger stripped off the colder ICM from the smaller subcluster. The diffuse radio emission obtained from LOFAR shows a remarkably similar horseshoe pattern found by \cite{Canning_2017} in Chandra X-ray images. Hence, it appears that there is a horseshoe-like pattern $\sim$ 100 kpc from the center along which there is turbulence. This causes heating of the ICM found in X-ray and diffuse emission in the radio along with the steepening of the spectral index along this direction.

\subsection{Abell 2261}
Abell 2261 is a relaxed galaxy cluster at z = 0.22 and is well studied in literature \citep{Umetsu_2009,Coe_2012,okabe_wl_subaru,von_der_Linden_2014, Spolaor_2017,Gultkein_2021}. \cite{Allen_2000} classified this cluster as a relaxed cool core cluster based on its X-ray properties. However, recent radio analysis by \cite{Sommer_2017} using VLA and \cite{Savini_2019} using LOFAR find a diffuse radio halo approximately 1Mpc in size. \cite{Sommer_2017} obtains an average spectral index of 1.2 while \cite{Savini_2019} obtained a value of 1.7. It has been suggested that disturbance/turbulence in clusters causes radio halos \citep{Van_weeren_2019}. It also appears there is a smaller X-ray clump of about 0.7 Mpc SW of the cluster. It is not clear if this clump is related to the main cluster. Another region of X-ray excess is found about 1 Mpc south of the cluster. A third region of X-ray excess is found to the NE of the cluster. X-ray images from XMM show these regions of extended emission. These are smaller groups and clusters. All correspond to an over-density of galaxies in our galaxy number density maps. Weak lensing analysis of this cluster has been studied by \cite{Umetsu_2009, okabe_wl_subaru, von_der_Linden_2014}. All studies recover the X-ray clump about 0.7 Mpc SW of the cluster. \cite{Umetsu_2009, okabe_wl_subaru} also recover a mass excess located 1Mpc south the cluster coincident to the X-ray excess. 

\begin{figure}[ht!]
    \centering
   
    \subfigure[]{\includegraphics[width=0.85\textwidth]{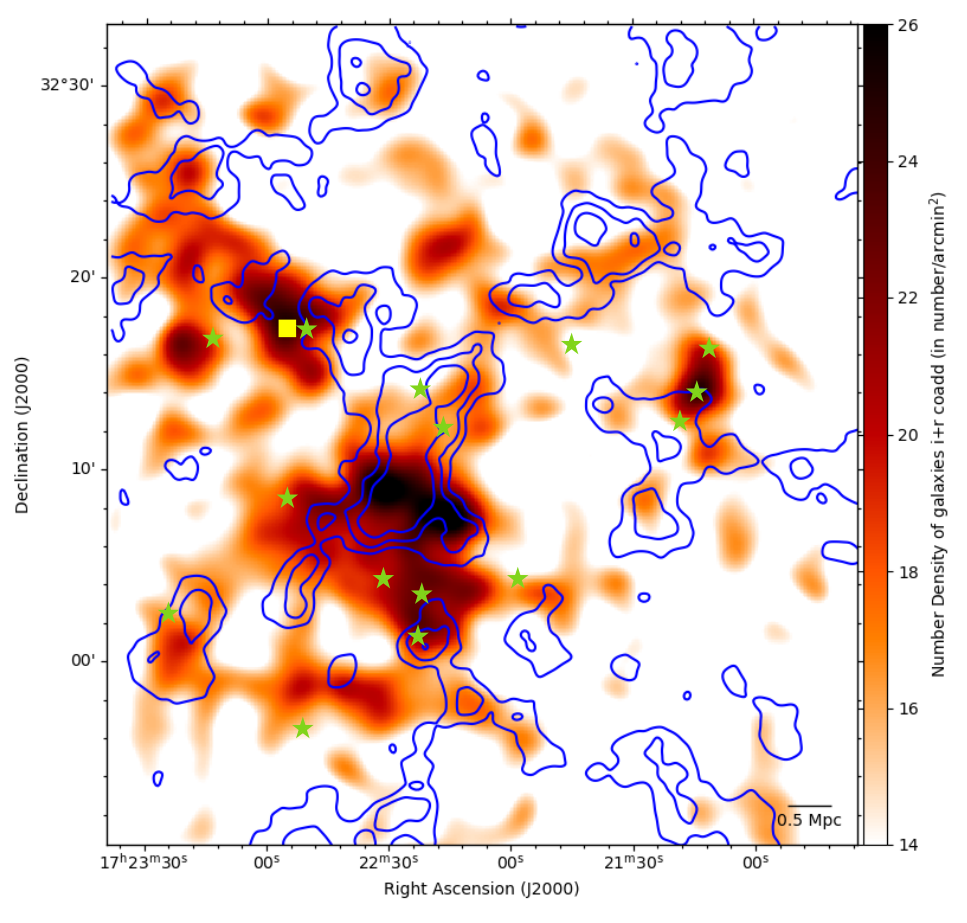}}
    
    \caption{ Contours of the E-Mode significance map of A2261 overlayed on the galaxy number density map of all sources except stars. The contours are drawn at 0.6, 1.5, and 2.5 standard deviations. The green stars show galaxy clusters and groups identified from NED using optical methods while yellow squares show clusters and groups obtained using X-ray. }
    \label{fig:A2261}
\end{figure}

\begin{figure}[ht!]
    \centering
    {\includegraphics[width=0.45\textwidth]{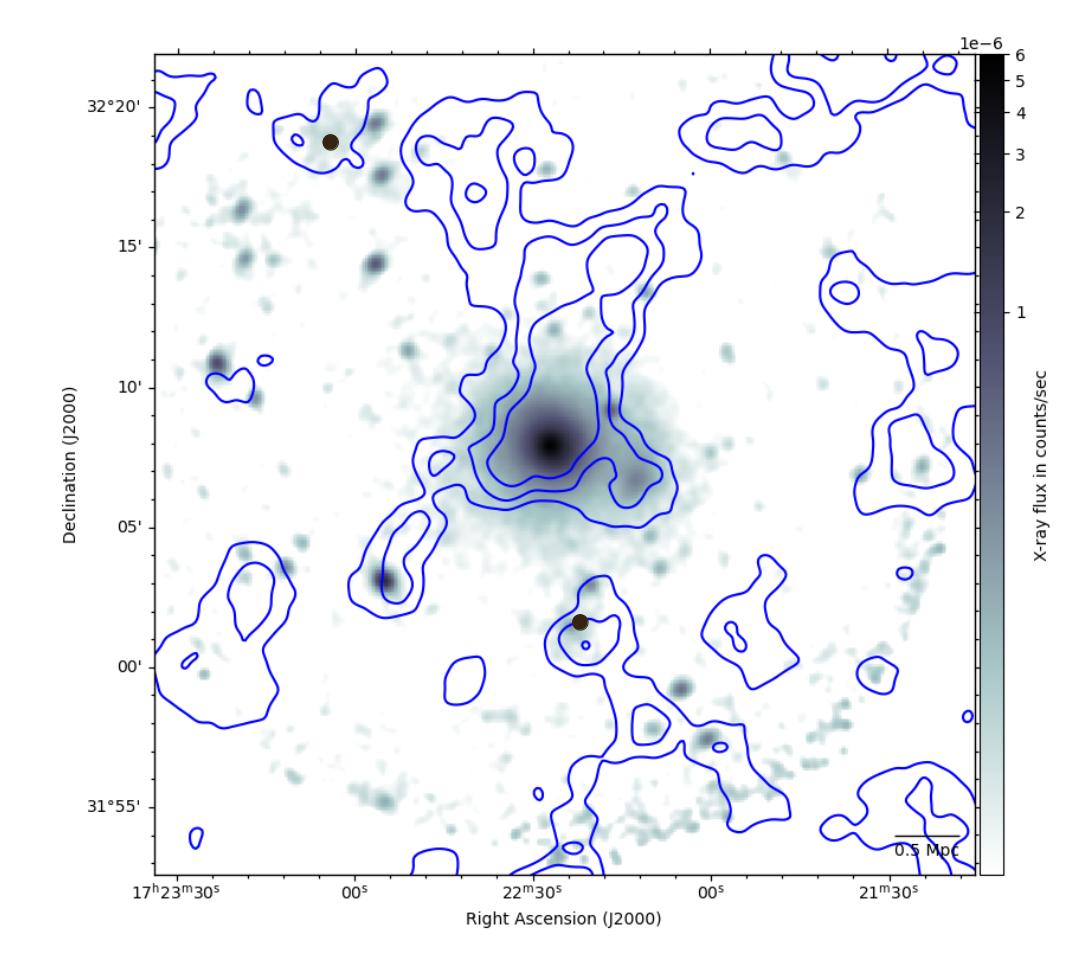}}
    {\includegraphics[width=0.45\textwidth]{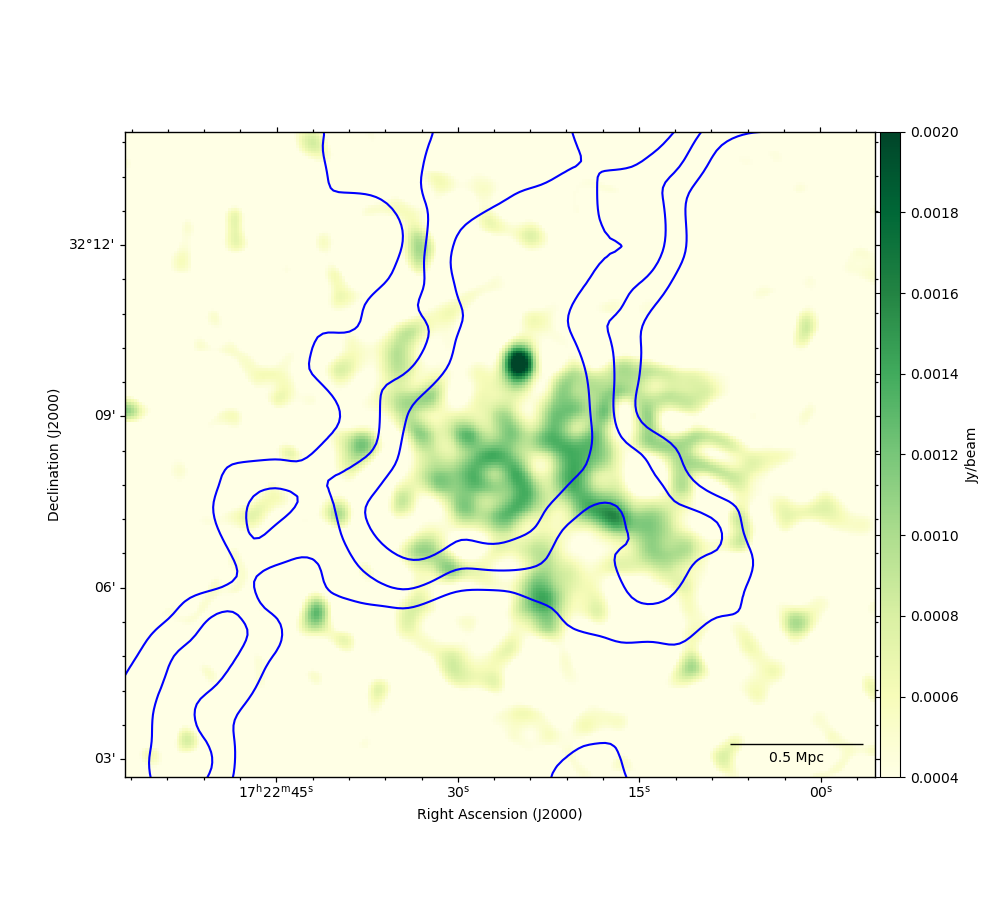}}
    \caption{ (a) Contours of the E-Mode significance map of A2261 overlayed on the XMM X-ray image. The solid black circles show the regions of diffuse X-ray emission. One such region is south of the cluster while another one is northeast of the cluster at the edge of the field. (b) The same contours are overlayed on the LOFAR image. }
    \label{fig:A2261_xray_radio}
\end{figure}

In our weak lensing analysis, we find the cluster along with some of the smaller groups and clusters are successfully recovered. In Figure \ref{fig:A2261} we overlay the significance contours of the E-Mode on the number density map of all sources except stars and point sources. The E-Mode contours have been drawn at 0.6, 1.5, and 2.5 standard deviations. The galaxy clusters and groups reported on NED are shown as green stars (optical methods) and yellow squares (X-ray methods). In Figure \ref{fig:A2261_xray_radio}a we show the contours of the mean E-mode significance map overlayed on the X-ray image from XMM. We use the XMM image instead of Chandra because of the wider field and more uniform coverage. This allows us to see the diffuse X-ray sources in the field which have been shown as black circles. In Figure \ref{fig:A2261_xray_radio}b we show the same contours overlayed on the radio image obtained from LOFAR. 

We find that the center of the cluster appears elongated in the NE-SW direction in optical number density maps. The radio image from LOFAR also shows a diffuse emission, especially in the SW direction. This makes the X-ray excess seen at approximately 0.7 Mpc SW of the cluster likely related to the cluster. This excess was also recovered in our significance map and is coincident with the X-ray excess as can be seen in Figure \ref{fig:A2261_xray_radio}a. We also find that some of the galaxy groups and clusters reported on NED are recovered using the mean E-Mode significance map as shown in Figure \ref{fig:A2261}. The recovery is worse than the case of A115 and comparable to A2219. The prominent galaxy over-density to the NW of the cluster (17h:21m:30, 32:15:00) is slightly recovered in the significance maps. The prominent cluster NE of the main cluster is recovered in the significance map. This smaller cluster can be seen as an extended X-ray source on the top left of Figure \ref{fig:A2261_xray_radio}a. To the SW of this cluster, we find another diffuse X-ray source which is marked by a black circle. This region also corresponds to a significant overdensity of galaxies and is recovered in our significance map. 

\begin{figure}[ht!]
    \centering
    {\includegraphics[width=0.45\textwidth]{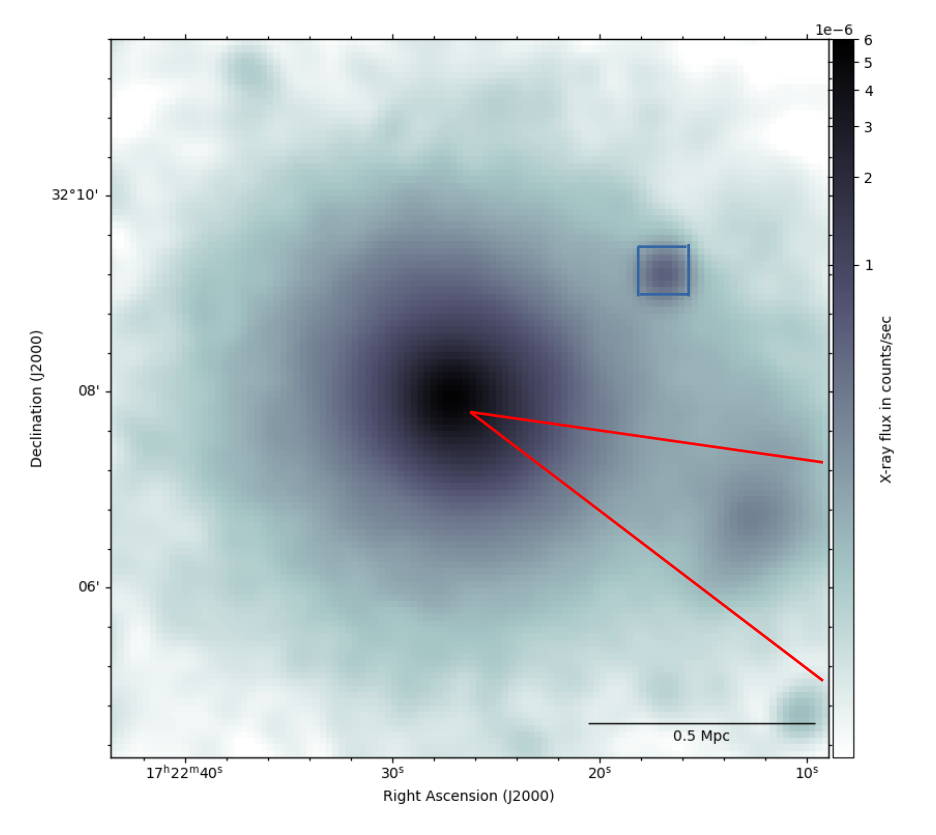}}
    {\includegraphics[width=0.45\textwidth]{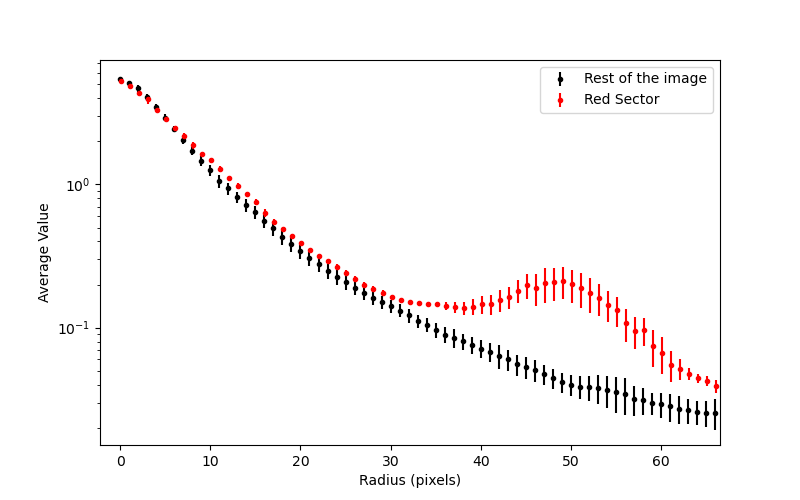}}
    \caption{ (a) The X-ray image of A2261 from XMM. The image is divided into 1-pixel wide radial bins and the sigma-clipped median and stated deviation is plotted in (b). The pixels in the blue square have been excluded from this analysis. Each pixel is 4$''$ in size i.e. 14.4 kpc. }
    \label{fig:A2261_xray_sector}
\end{figure}

In order to investigate the X-ray clump 0.7 Mpc SW of the cluster we perform radial binning on the X-ray image. Fitting an elliptical $\beta$ profile to the main cluster, it was found to be almost perfectly circular. We use a 40 x 40 pixel cutout centered on the central pixel. This corresponds to a size of 0.58 Mpc. This region is chosen to minimize the effect of the bright stars and the smaller clump on the fit. The best-fit value of $\beta $ is 0.55 while the characteristic radius in the x and y axes are 93.2 and 94.2 kpc respectively. The fitted parameters are insensitive to the cutout size used for the fit. These values are consistent with what is expected from relaxed clusters \citep{vikh_2006, kafer_2019}. Since the best-fit profile is almost exactly circular, we divide the X-ray image from XMM shown in Figure \ref{fig:A2261_xray_sector}a into radial bins of 1 pixel wide. We divide these circular bins into two sectors defined by the red lines. We also reject any pixels in the blue square to avoid contamination from the star. We then compute the sigma-clipped median and standard deviation of all pixels in the two sectors and show the results in Figure \ref{fig:A2261_xray_sector}b. The red points always lie slightly higher than the black points. While the difference is not statistically significant, it favors the theory that a smaller clump is likely related to the main cluster and is interacting with it. The X-ray emission and the radio emission (without compact component subtracted) from this clump appear to coincide with a giant elliptical galaxy in our images. The photometric redshift of the elliptical galaxy could not be determined with high reliability due to blending with two nearby sources. However, a conservative visual estimate of the galaxy's size, assuming it is located at the cluster redshift, is over 40 kpc. Hence, it is likely that this giant elliptical galaxy is at cluster redshift or closer. 

\subsection{Conclusion}
In this paper, we have presented a weak lensing analysis of Abell 115, Abell 2219, and Abell 2261 using short exposures obtained from WIYN-ODI. The overall depth of the co-added image is 27th magnitude for Abell 115 and 26th magnitude for Abell 2261 and Abell 2219. We find that using mean $M_{ap}$ over variance maps helps us recover most of the mass structures in our field of view, so long as we have good-quality optical data. This becomes evident for A2261 and A2291 for which we have the worse quality data which leads to significantly worse results. For Abell 115 we find separation of DM and the hot ICM gas. This is not unexpected since A115 is undergoing a merger. However, the galaxy number density and the lensing contours appear largely consistent. This is consistent with individual galaxies and DM behaving collisionless in a merger. We find the most likely explanation is that the northern clump is moving into the plane while the southern clump is moving out of the plane. This is supported by spectroscopic studies. The long filamentary structure connecting the northern sub-clump of A115 is also recovered. The radio relic in this cluster lines up perfectly with our E-Mode contours. We are also able to recover an almost 10Mpc long filamentary structure inside which A2219 is located. We find that it is likely that A2219 is accreting material along this structure from both directions. This is confirmed by the multiple shocks found in this system using deep Chandra image. Remarkably, both the radio image and X-ray image show a distinct horseshoe shape indicating an ongoing merger from the SW side. Finally, for A2261, we recovered all the regions showing diffuse X-rays using the E-Mode contours. While A2261 appears to be an extremely relaxed cluster with $\beta$ = 0.55 and $r_c \sim$ 100 kpc, radio analysis suggests ongoing interaction with a clump approximately 700kpc SW of the cluster, likely an infalling cluster. A radio analysis using VLA and LOFAR data suggests that A2261 has a very high spectral index of -1.7. Radio analysis suggests that both A2261 and A2219 have a disturbed core and hence produce diffuse radio emission over scales of 1Mpc.  
\section{Acknowledgments}
We thank the anonymous referee for their helpful comments. The authors would like to thank Purdue University for its continued support. We are also very grateful to the WIYN-ODI PPA team, especially Wilson Liu, Nick Smith, and all the telescope operators for their help. We thank Annalisa Bonafede and Shubham Bhagat for their help in interpreting the radio data. We also thank the Purdue Rosen Center for Advanced Computing (RCAC) for access to computing facilities that have been extensively used in this paper. This data analysis has been done on python and the authors acknowledge the use of astropy\citep{astropy:2013, astropy:2018, astropy:2022}, numpy\citep{harris2020array}, scipy\citep{2020SciPy-NMeth}, matplotlib\citep{Hunter:2007} and aplpy \citep{aplpy2012, aplpy2019}.This research has made use of the NASA/IPAC Extragalactic Database (NED), which is funded by the National Aeronautics and Space Administration and operated by the California Institute of Technology.

\bibliography{sample631}{}
\bibliographystyle{aasjournal}



\end{document}